\documentclass{iaus260}
\usepackage{graphicx}
\usepackage{url}

\title[Astronomy and astronomers in Jules Verne's novels] 
{Astronomy and astronomers\\ in Jules Verne's novels}

\author[Jacques Crovisier]   
{Jacques Crovisier}

\affiliation{Observatoire de Paris, \\ 5 place Jules Janssen, 92195 Meudon,
France \\email: {\tt jacques.crovisier@obspm.fr}}

\pubyear{2009}
\volume{260}  
\pagerange{119--126}
\jname{The R\^ole of Astronomy in Society and Culture}
\editors{D. Valls-Gabaud \& A. Boksenberg, eds.}

\begin{document}

\maketitle

\begin{abstract}
Almost all the \textit{Voyages Extraordinaires} written by Jules Verne refer to
astronomy.  In some of them, astronomy is even the leading theme.  However,
Jules Verne was basically not learned in science.  His knowledge of astronomy
came from contemporaneous popular publications and discussions with specialists
among his friends or his family.  In this article, I examine, from the text and
illustrations of his novels, how astronomy was perceived and conveyed by Jules
Verne, with errors and limitations on the one hand, with great respect and
enthusiasm on the other hand.  This informs us on how astronomy was understood
by an ``honn\^ete homme'' in the late 19th century.

\end{abstract}

\firstsection 
\section{Introduction}

\begin{figure}
\begin{center}
\includegraphics[width=7cm]{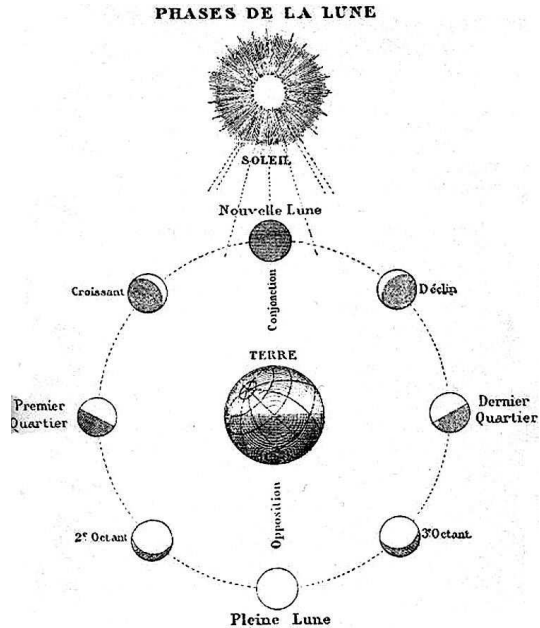} 
\caption{The phases of the Moon, a didactic illustration of \textit{From the
Earth to the Moon}.}
   \label{fig1}
\end{center}
\end{figure}

Jules Verne (1828--1905) wrote more than 60 novels which constitute the
\textit{Voyages Extraordinaires} series\footnote{An authoritative biography of
Jules Verne in English was recently published by Butcher (\cite{butc_2006}).}.
Most of them were scientific novels, announcing modern science fiction.
However, following the strong suggestions of his editor Pierre-Jules Hetzel,
Jules Verne promoted science in his novels, so that they could be sold as
educational material to the youth (Fig.~\ref{fig1}).  Jules Verne had no
scientific education.  He relied on popular publications and discussions with
specialists chosen among friends and relatives.

This article briefly presents several examples of how astronomy appears in the
text and illustrations of the \textit{Voyages Extraordinaires}.  \sloppy Further
information can be found in Bacchus (\cite{bacc_1992}), Crovisier
(\cite{crov_2005a}), Sauzereau \& Giton (\cite{sauz-gito_2006}) and
\url{http://www.lesia.obspm.fr/perso/jacques-crovisier/JV/verne_gene_eng.html}

\section{Astronomical voyages}

Several novels of Jules Verne explicitly refer to astronomy (Table~1).  In them,
the story line is occasionally interrupted by full pages, or even chapters,
devoted to didactic explanations of astronomical topics.  In addition, almost
all other novels of the series contain episodic references to astronomy or
connected sciences.  In particular, determining the ship's bearing is a
recurrent affair.  Jules Verne was an amateur sailor himself (he successively
owned three yachts).  His brother Paul (1829--1897) had a certificate of master
mariner and advised him for navigation matters.

\bigskip

\noindent \textbf{\textit{De la Terre \`a la Lune}} (\textit{From the Earth to
the Moon}) and \textbf{\textit{Autour de la Lune}} (\textit{Around the Moon})
are probably the most famous novels on lunar exploration.  Jules Verne required
the advice of his cousin Henri Garcet (1815--1871), a professor of mathematics
in Paris who published \textit{Le\c{c}ons nouvelles de cosmographie} (Garcet
1854), a textbook on astronomy which was popular for several decades.  Garcet's
former colleague Joseph Bertrand (1822--1900), a distinguished mathematician and
academician, also helped.

In \textit{Around the Moon}, the projectile on its way to the Moon flies by a
meteor, which is identified as the ``second satellite of the Earth'' of M.
Petit.  Indeed, Fr\'ed\'eric Petit (1810--1865), founder and director of
Toulouse Observatory, studied the orbits of meteors and suggested that one of
them could be a second satellite of the Earth (Kragh \cite{krag_2009}). This
hypothesis was soon abandoned due to the uncertainty on the determination of the
orbit of such bodies.  Jules Verne probably did not read the original reports of
Petit (published in the \textit{Comptes-rendus}), but took the information from
the popular book of Am\'ed\'ee Guillemin (1826--1893) on the Moon (Guillemin
\cite{guil_1866}).

\bigskip

\begin{figure}[h]
\begin{center}
\includegraphics[width=6.00cm]{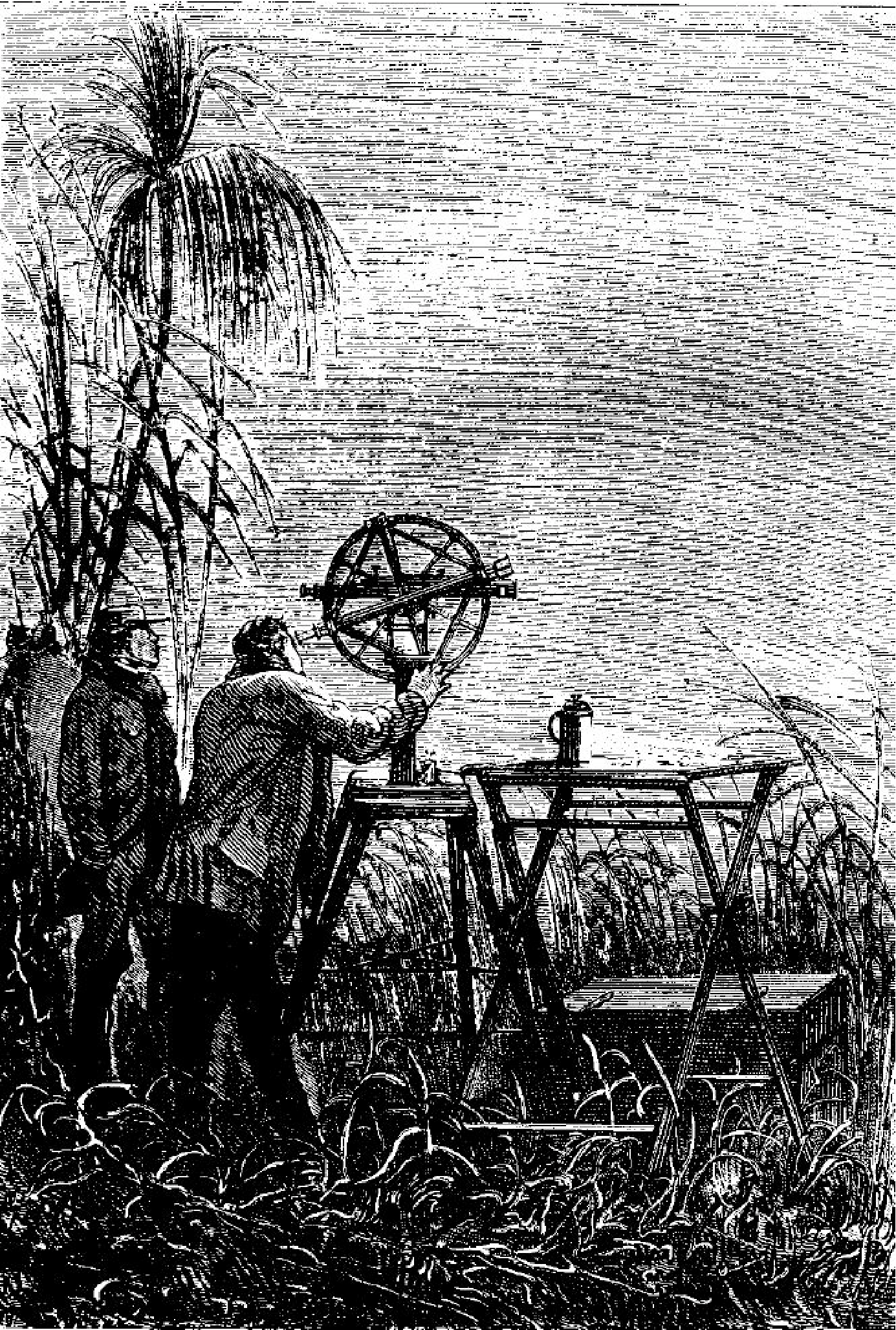} 
\hspace{0.5cm}
\includegraphics[width=5.60cm]{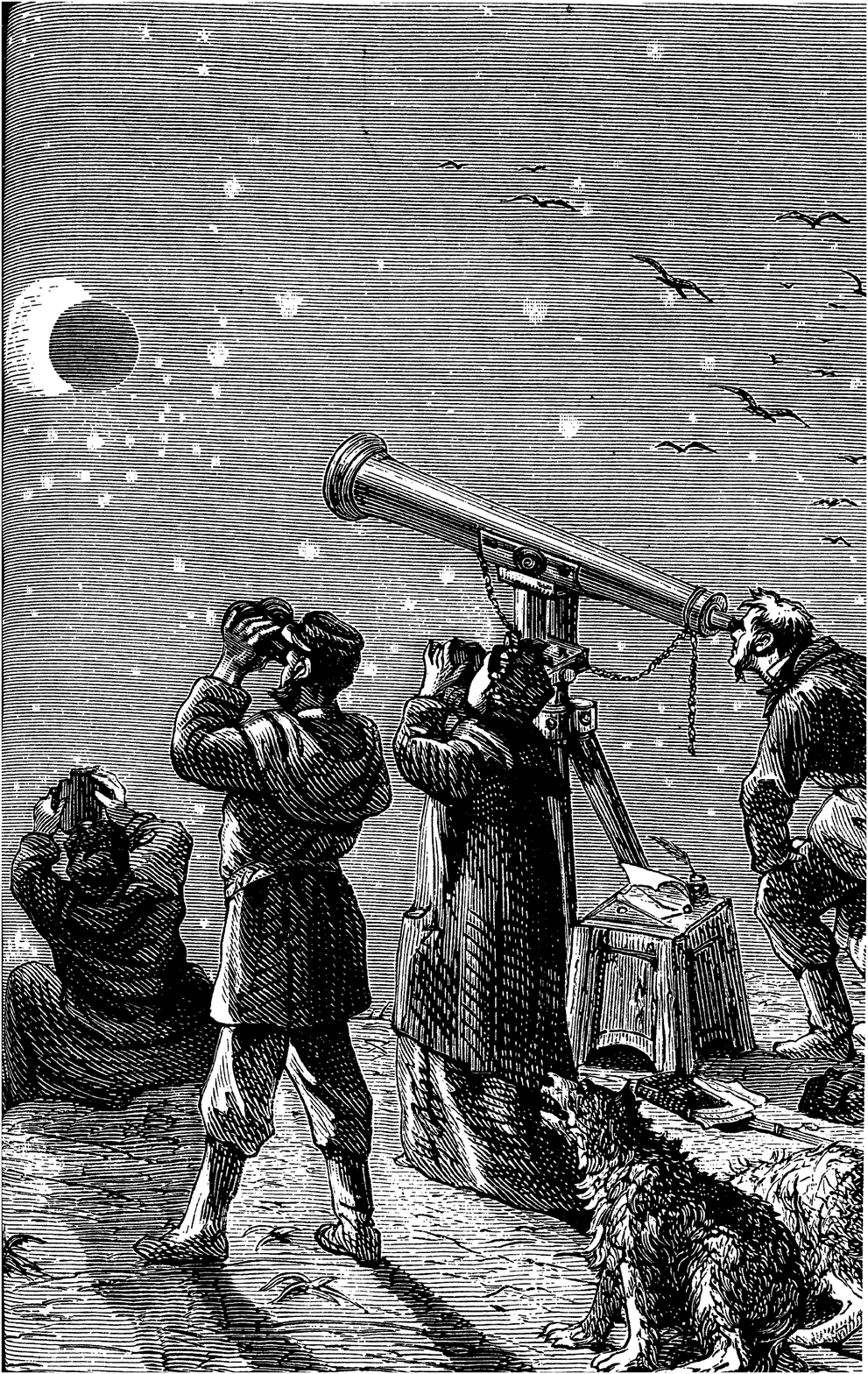} 
\caption{Astronomers at work from illustrations of Jules Verne's novels.  Left (a):
Measurement of an arc of meridian (drawing by J. F\'erat, from
\textit{Adventures of three Russians and three Britons in Southern Africa"}).
Right (b): Observation of a solar eclipse in the American far North (drawing by J.
F\'erat, from \textit{The Fur Country}).}
   \label{fig2}
\end{center}
\end{figure}

\noindent \textbf{\textit{Aventures de trois Russes et de trois Anglais dans
l'Afrique australe}} (\textit{Adventures of Three Russians and Three Britons in
Southern Africa}) transposes in Africa the great astronomical expeditions to
measure arcs of meridian (Fig.~\ref{fig2}a).  A multinational team of
astronomers strive to accomplish their mission in hostile conditions.  This is a
tribute to Fran\c{c}ois Arago.  Jules Verne also acknowledges his cousin H.
Garcet, by reproducing a page of his \textit{Cosmographie} explaining
triangulation.

\bigskip

\noindent \textbf{\textit{Le Pays des Fourrures}} (\textit{The Fur Country})
relates the failed observation of the 18 July 1860 solar eclipse in the American
far North (Fig.~\ref{fig2}b).  Several expeditions were indeed organized to
observe this eclipse, especially in Spain and Africa, but the eclipse was not
total in \textit{The Fur Country}!

\bigskip 

\noindent \textbf{\textit{Le Tour du monde en quatre-vingt jours}}
(\textit{Around the World in Eighty Days}), which deals with the oddities of the
change of date when travelling around the world, is presented as a
``cosmographic joke'' by J. Verne himself.  It was prompted to him by Edgar~A.
Poe's (1809--1849) short story \textit{Three Sundays in a Week}.  (However, this
curiosity was already remarked in 1524 by the companions of Magellan in their
diary.)  This story is at the origin of a talk on meridians and calendars
at the Soci\'et\'e g\'eographique in Paris, resulting to the only scientific
publication by Verne (\cite{vern_1873}).

\bigskip 

\noindent \textbf{\textit{Le Rayon-vert}} (\textit{The Green Flash}) is an
exemplary case of the synergy between science and Jules Verne.  We still ignore
when, where and how Jules Verne learned of the green flash (O'Connell,
\cite{ocon_1958}).  But its report in this novel\footnote{It is also mentioned
in four other novels of the \textit{Voyages Extraordinaires} series.} indeed
triggered research on this atmospheric phenomenon, previously unknown to the
layman!

\begin{table}
\begin{center}
\caption{Novels of Jules Verne with references to astronomy.  The date is that
of the first Hetzel edition.  Many variants exist for the English titles.}
\begin{tabular}{rcl}
\hline
\multicolumn {3}{c}{\textit{\textbf{strong references}}}\\
De la Terre \`a la Lune  & 1865 & From the Earth to the Moon \\
Autour de la Lune        & 1870 & Around the Moon \\
Aventures de trois Russes et de trois  & 1872 & Adventures of Three Russians and Three \\
Anglais dans l'Afrique australe        &      & Britons in Southern Africa\\
Le Pays des Fourrures    & 1872 & The Fur Country \\
Hector Servadac          & 1877 & Hector Servadac (Off on a Comet)\\
Sans dessus dessous      & 1889 & Topsy-Turvy \\
La Chasse au m\'et\'eore & 1908 & The Chase of the Golden Meteor\\
\hline
\multicolumn {3}{c}{\textit{\textbf{secondary references}}}\\
Voyages et aventures du capitaine Hatteras & 1866 & The Adventures of Captain Hatteras \\
Le Tour du monde en quatre-vingt jours & 1873 & Around the World in Eighty  Days \\
Les Cinq cents millions de la B\'egum  & 1879 & The Begum's Millions \\
Le Rayon-vert            & 1882 & The Green Ray \\
Mirifiques aventures de Ma\^itre Antifer & 1894 & Adventures of Captain Antifer \\
L'\^{I}le \`a h\'elice   & 1895 & Propellor Island \\
\hline
\multicolumn {3}{c}{\textit{almost all other novels contain episodic references to}}\\
\multicolumn {3}{c}{\textit{astronomy, geography, 
geodesy, hydrography, meteorology\ldots}}\\
\hline
\end{tabular}\textasciitilde
\end{center}
\end{table}

\bigskip 

\begin{figure}[h]
\begin{center}
\includegraphics[width=5.56cm]{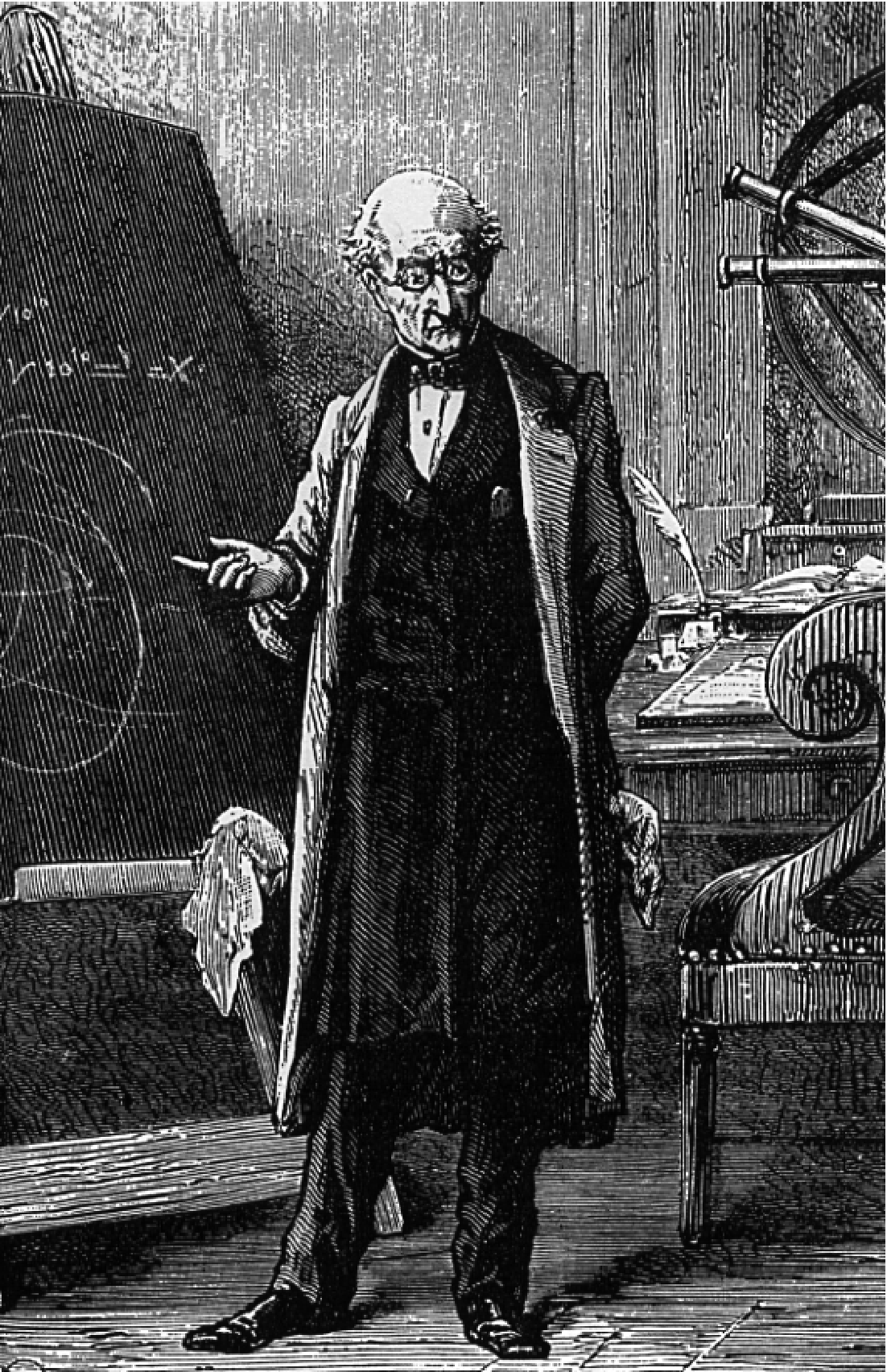} 
\hspace{0.5cm}
\includegraphics[width=6.00cm]{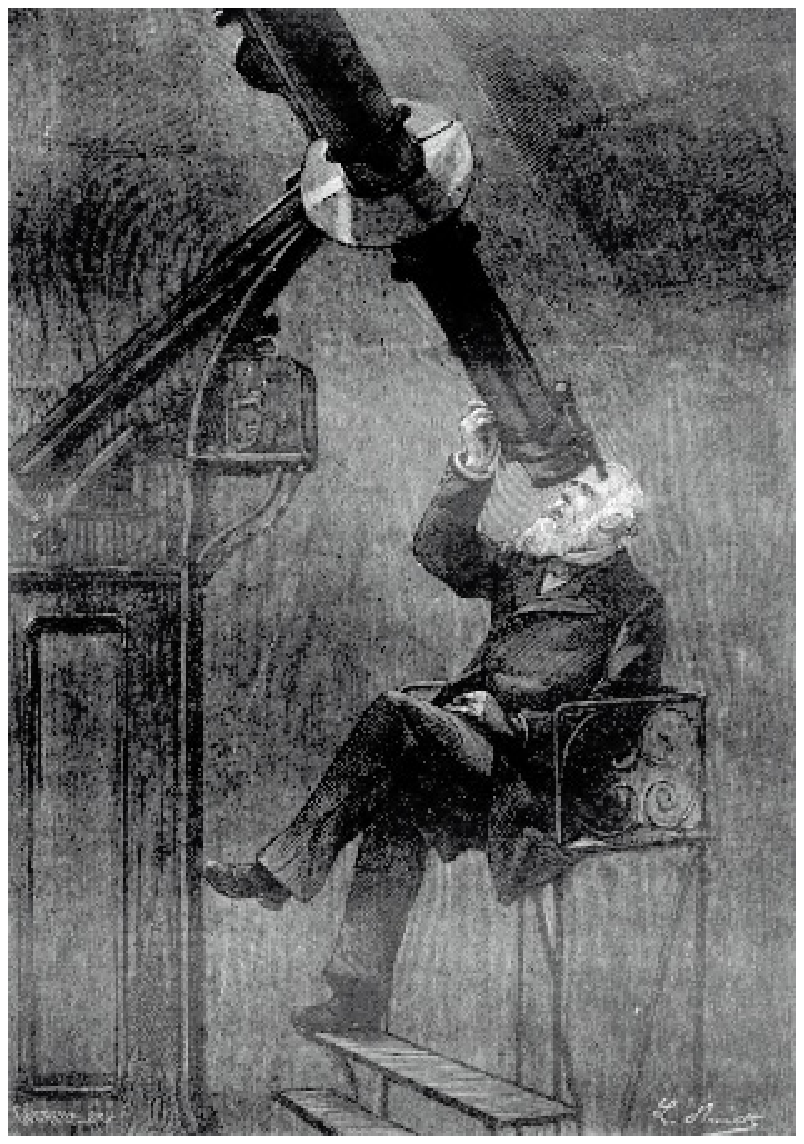} 
\caption{Portraits of astronomers in Jules Verne's novels.  Left (a): An irascible
astronomer, Palmyrin Rosette (drawing by P. Philippoteaux, from \textit{Hector
Servadac}).  Right (b): A kind astronomer, the former king of Mal\'ecarlie, modelled
after the emperor of Brazil Dom Pedro II (drawing by L. Benett, from
\textit{L'\^{I}le \`a h\'elice} (\textit{Propellor Island})).}
   \label{fig3}
\end{center}
\end{figure}

\noindent \textbf{\textit{Hector Servadac}} tells the voyage across the Solar
System \textit{aboard} a comet of a small community, including the free-lance
astronomer Palmyrin Rosette (Fig.~\ref{fig3}a).  It prefigures the Voyager
missions of the NASA which flew-by successively several planets, as well as the
ongoing Rosetta mission of ESA towards comet 67P/Churyumov-Gerasimenko
(Crovisier, \cite{crov_2005b}).

When working on this novel, Jules Verne reduced the orbital period of the comet
--- which crossed the orbits of Venus and Jupiter --- to two years, in obvious
contradiction with Kepler's laws.

\bigskip 

\noindent \textbf{\textit{Sans dessus dessous}} (\textit{Topsy-Turvy}) is a
foolish attempt to tilt the rotation axis of the Earth, using the recoil effect
of a giant cannon.  The attempt fails, due to a miscalculation.  The story is
based upon a sound technical study (Badoureau \cite{bado_2005}).

Albert Badoureau (1853--1923), a mining engineer, lived in Amiens from 1884 to
1894 where he was meeting Jules Verne at the Soci\'et\'e industrielle and the
Acad\'emie.  He was commissioned (and paid) by Jules Verne to provide
the scientific background of \textit{Topsy-Turvy}.  Indeed, one of the
characters of this novel --- Alcide Pierdeux --- is sketched after Badoureau.
The study of Badoureau was published in extenso, packed with formulas and
scientific drawings, as the last chapter in the first edition of the novel.

\bigskip 

\noindent \textbf{\textit{La Chasse au m\'et\'eore}} (\textit{The Chase of the
Golden Meteor}), a posthumous novel rewritten by Jules Verne's son Michel
(1861--1925), narrates the rivalry between two amateur astronomers who both
discovered a bolid (in fact an asteroid).  The asteroid (like the comet in
\textit{Hector Servadac}) is made of gold, and the announcement of its fall on
Earth provokes a financial crisis.

One can remark that the orbital elements of the asteroid (both those given in
the original Jules' version or the rewritten Michel's version) do not fit with
Kepler's laws.  That was also the case for the bolid of \textit{Around the
Moon}.

One of the additions of Michel Verne to the original text of his father was the
introduction of a new character.  Z\'ephyrin Xirdal is an absent-minded
scientist, not unlike Alcide Pierdeux in \textit{Topsy-Turvy}.  He used a device
based on the equivalence between mass and energy to deviate the orbit of the
asteroid.  Michel had probably heard of the new ideas of Einstein, which were
emerging at that time (1907).

\begin{figure}
\begin{center}
\includegraphics[width=6.0cm]{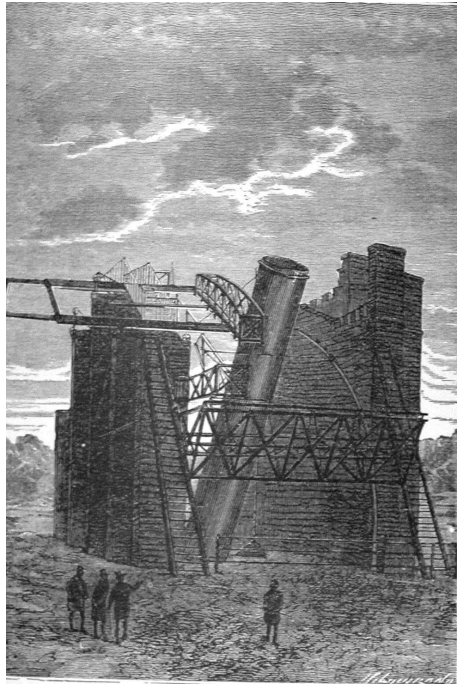} 
\caption{The 16-feet telescope of Long's Peak observatory, in \textit{Around the
Moon}, was designed as an enlargement of the 6-feet telescope of Lord Rosse
(William Parsons), the largest telescope in the world at that time.  It
was placed atop a mountain, as are now modern observatories.}
   \label{fig4}
\end{center}
\end{figure}

\section{Conclusion}

Verne had a very high opinion of astronomy and astronomers.  For him, ``an
astronomer is more than human, since he lives away from the
Earth''\footnote{``Un astronome est plus qu'un homme, puisqu'il vit en dehors du
monde terrestre'' (\textit{Hector Servadac}, part II, chap.  XIX).}.
Fran\c{c}ois Arago was his idol (Le Lay, \cite{lela_2005}).  Astronomers are
described in his novels as proficient professionals, somewhat between
absent-minded scientists and mad scientists (Fig.~\ref{fig3}).

There are basically no inventions in Verne's novels.  Fantasy is not present;
for instance, the controversial question of the existence of \textit{Martians}
is carefully avoided (which is not the case in the popular books of Camille
Flammarion (1842--1925), Verne's rival in literature).

Jules Verne extrapolated from existing techniques and machines.  He na\"{i}vely
assumed that instrumental performances linearly scale as the size (as for the
telescope in \textit{Around the Moon}; Fig.~\ref{fig4}), or that the accuracy of
measurements can be infinitely improved by the repetition of observations (as in
\textit{Adventures of Three Russians and Three Britons in Southern Africa}).

Some of the ideas of Jules Verne are still quite pertinent today.  For instance,
in \textit{The Adventures of Captain Hatteras}, Dr Clawbonny --- an eclectic
character who discourses on various scientific topics for the benefit of the
reader --- tells:

\begin{quote}
    
Comets are the deus ex machina; every time you're embarrassed in cosmography,
you bring in a comet.  It is the most helpful heavenly body I know, for at the
slightest sign from scientists, it does its level best to fix
everything.\footnote{``La com\`ete est le \textit{Deus ex machina}~; toutes les
fois qu'on est embarrass\'e en cosmographie, on appelle une com\`ete \`a son
secours.  C'est l'astre le plus complaisant que je connaisse, et, au moindre
signe d'un savant, il se d\'erange pour tout arranger~!''  (\textit{Voyages et
aventures du capitaine Hatteras}, 1866, Part 2, Chap.  14).}

\end{quote}

\bigskip

\noindent Jules Verne had in mind the changes of the spin axis of the Earth
induced by cometary impacts on the Earth, leading to climatic changes.  He would
be amused to learn that still nowadays, comets are invoked to explain the origin
of water on Earth, the emergence of life, or the queer properties of
circumstellar discs.



\end{document}